\documentclass[pre,amssymb,amsmath,floats,twocolumn,showpacs,superscriptaddress]{revtex4}

\usepackage{epsfig}

\begin{document}

\title{BKT-like transition in the Potts model on an inhomogeneous annealed network}

\author{E. Khajeh}
\email{e-khajeh@fis.ua.pt}
\affiliation{Departamento de F{\'\i}sica da Universidade de Aveiro, 3810-193 Aveiro, Portugal}
\affiliation{Department of Physics, Shahid Beheshti University, Evin, Tehran, 19839, Iran}

\author{S. N. Dorogovtsev}
\email{sdorogov@fis.ua.pt}
\affiliation{Departamento de F{\'\i}sica da Universidade de Aveiro, 3810-193 Aveiro, Portugal}
\affiliation{A. F. Ioffe Physico-Technical Institute, 194021
  St. Petersburg, Russia} 

\author{J. F. F. Mendes}
\email{jfmendes@fis.ua.pt}
\affiliation{Departamento de F{\'\i}sica da Universidade de Aveiro, 3810-193 Aveiro, Portugal}

\begin{abstract} 
We solve the ferromagnetic $q$-state Potts model on an inhomogeneous annealed network which mimics a random recursive graph. We find that this system has the inverted Berezinskii--Kosterlitz--Thouless (BKT) phase transition for any $q \geq 1$, including the values $q \geq 3$, where the Potts model normally shows a first order phase transition. We obtain the temperature dependences of the order parameter, specific heat, and susceptibility demonstrating features typical for the BKT transition. We show that in the entire normal phase, both the distribution of a linear response to an applied local field and the distribution of spin-spin correlations have a critical, i.e. power-law, form.        
\end{abstract}

\pacs{05.50.+q, 05.10.-a, 05.40.-a, 87.18.Sn}

\maketitle


\section{Introduction}\label{s-introduction}


Among numerous unusual phenomena which were observed in complex networks \cite{ab02,dm02,n03,pvbook04,blm06}, one effect is 
astonishing. 
In a wide range of growing networks, a phase transition of the birth of the giant connected component demonstrates the Berezinskii-Kosterlitz-Thouless (BKT) singularity \cite{chk01,dms01,l02,kkkr02,bb03,cb03,d03,br05,kd04}, i.e., the size $S$ of the giant connected component has the critical singularity $S \propto \exp{(-\text{const}/\sqrt{\delta})}$. Here the appropriate parameter $\delta$ characterizes the deviation from the critical point. Recall that in cooperative models on lattices, an (infinite order) BKT transition is possible in very specific situations (e.g., in the two-dimensional $XY$ model) where critical fluctuations are quite strong \cite{b71,kt73}, on the lower critical dimension. In contrast, networks are extremely compact, infinite-dimensional objects (small worlds) and so can be described by, in essence, mean field theories. Consequently the reason for the BKT singularity in growing network systems strongly differs from that for the BKT transition in solid state systems. It was argued that it is the inhomogeneous architecture of some growing networks that results in the BKT singularities. Note that in equilibrium networks with any degree distributions, where all vertices are statistically equivalent, this transition is impossible \cite{pv01,cbh02,ahs02,dgm02,lvv02,b02,it02,dgm04,em05,ghi06}. 

Interestingly, a similar critical behaviour was earlier (already in 1990!) found, proved, and analysed in the ferromagnetic Ising model for a chain of spins with a specific large-scale inhomogeneity of long-range interactions \cite{ccg90,cc91}. (In Ref.~\cite{cc91}, the BKT singularity was also revealed in another cooperative model on the same substrate, where, instead of the Ising spins, there were continuous scalar fields.) In a network context, actually the same system was studied in Ref.~\cite{bcd05}, where a random recursive graph (i.e., with quenched disorder) was approximately substituted by a graph with annealed disorder---{\em an annealed network}. (By definition, in the random recursive graph, at each time step, a new vertex is attached to a randomly chosen existing vertex.) The resulting model may be analysed exactly, and it is equivalent to the Ising model on a deterministic fully connected graph with inhomogeneous Ising interactions between spins (see Fig.~\ref{f1}). The BKT singularity in the Ising model on another, hierarchically organized, deterministic network were found in Ref.~\cite{hb06}. 

One should stress that on inhomogeneous networks, the BKT transition was found in the cooperative models (the Ising model, percolation) which have a second order phase transition in high-dimensional lattices and the 
classical random graphs. 
In contrast, in the present paper we 
solve 
the $q$-state ferromagnetic Potts model, which has a first order phase transition on the classical random graphs if $q \geq 3$ \cite{w82}. 
That is, there is a jump of the order parameter and hysteresis if the substrate 
of the model is an infinite-dimensional lattice. We show that even in this, quite different situation, a BKT-like infinite-order phase transition is realised on the annealed variation of the random recursive graph, and the jump and hysteresis are absent. 
We calculate a set of observables and find typical BKT singularities. Furthermore, we obtain distribution of the linear response to local field---the distribution of ``correlation volume''---and find that it has a power law form in the entire normal phase. 
Recall that in cooperative models on equilibrium networks, e.g., on uncorrelated networks, this correlation characteristic is a power law only at a critical point.  
We also obtain the distribution of pair spin-spin correlations, which was recently introduced in Ref.~\cite{mr05}, and again find a power law in the entire normal phase. Thus we generalize results obtained for the Ising model on this network to the case of the Potts model with an arbitrary number of states. 
The revealed critical behavior dramatically differs from that of the Potts model on the configuration model \cite{dgm04}.  
Technically, our derivation directly follows Ref.~\cite{bcd05}.      

This paper is organized as follows. In Sec.~\ref{s-model}, we introduce the model, in Sec.~\ref{s-results}, we present our main results, and in the following sections derive and discuss them in detail.


\section{The model}\label{s-model} 


The Hamiltonian of the ferromagnetic $q$-state Potts model is 
\begin{equation}
\mathcal{H}=-
{\displaystyle\sum\limits_{\left\langle ij\right\rangle }}
\delta(s_{i},s_{j})-%
{\displaystyle\sum\limits_{i=0}^{t}}
H_{i}\delta(s_{i},1), 
\label{.5}
\end{equation}
where the coupling is set equal to $1$, $\left\langle ij\right\rangle $ indicates summing over the coupled spins, and $s_{i}=1,2,...,q$. $\delta(s_{i},s_{j})$ is the Kronecker delta symbol, and $H_{i}$ is the external
field. At $q=1$, this model is equivalent to the site percolation problem \cite{kf69}, and at $q=2$, it is the Ising model. For these two values of $q$, at the standard mean-field regime, the Potts model have a second order phase transition. For $q \geq 3$, the standard mean-field theory gives the first order phase transition.  

\begin{figure}[ptb]
\begin{center}
\scalebox{0.22}{\includegraphics[angle=0]{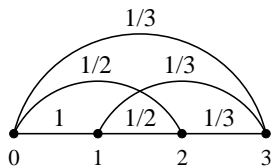}}
\end{center}
\caption{The deterministic fully connected graph which is equivalent to the 
asymmetric annealed network under study. 
Here $t=3$. The values of the Potts
couplings are shown on the edges.}
\label{f1}%
\end{figure}

We place the Potts model on the following asymmetric annealed network. Vertices are labelled $i=0,1,2,\ldots,t$, as in a growing network. Each vertex, say vertex $i$, have a single link of unit strength to previous vertices $j=0,1,2,\ldots,i-1$. The second end of this connection frequently hops at random among $j=0,1,2,\ldots,i-1$, which just means specific asymmetric annealing. The resulting network is equivalent to the fully connected graph with a large-scale inhomogeneity of the coupling (see Fig.~\ref{f1}). The mean-field theory for this network is exact, and it also mimics the more difficult case of the random recursive graph---the case of a quenched disorder \cite{remark}. 

In other words, in this model, the spin which was born at time $i$, has equal coupling $1/i$ to each of the older spins. 
That is, the Hamiltonian of our system is     
\begin{equation}
\mathcal{H}=-%
{\displaystyle\sum\limits_{0\leq i<j\leq t}}
\frac{\delta(s_{i},s_{j})}{j}-%
{\displaystyle\sum\limits_{i=0}^{t}}
H_{i}\delta(s_{i},1). 
\label{1}%
\end{equation}


\section{Main results}\label{s-results} 


\begin{figure}[ptb]
\epsfxsize=61mm
\begin{center}
\scalebox{0.1372}{\includegraphics[angle=0]{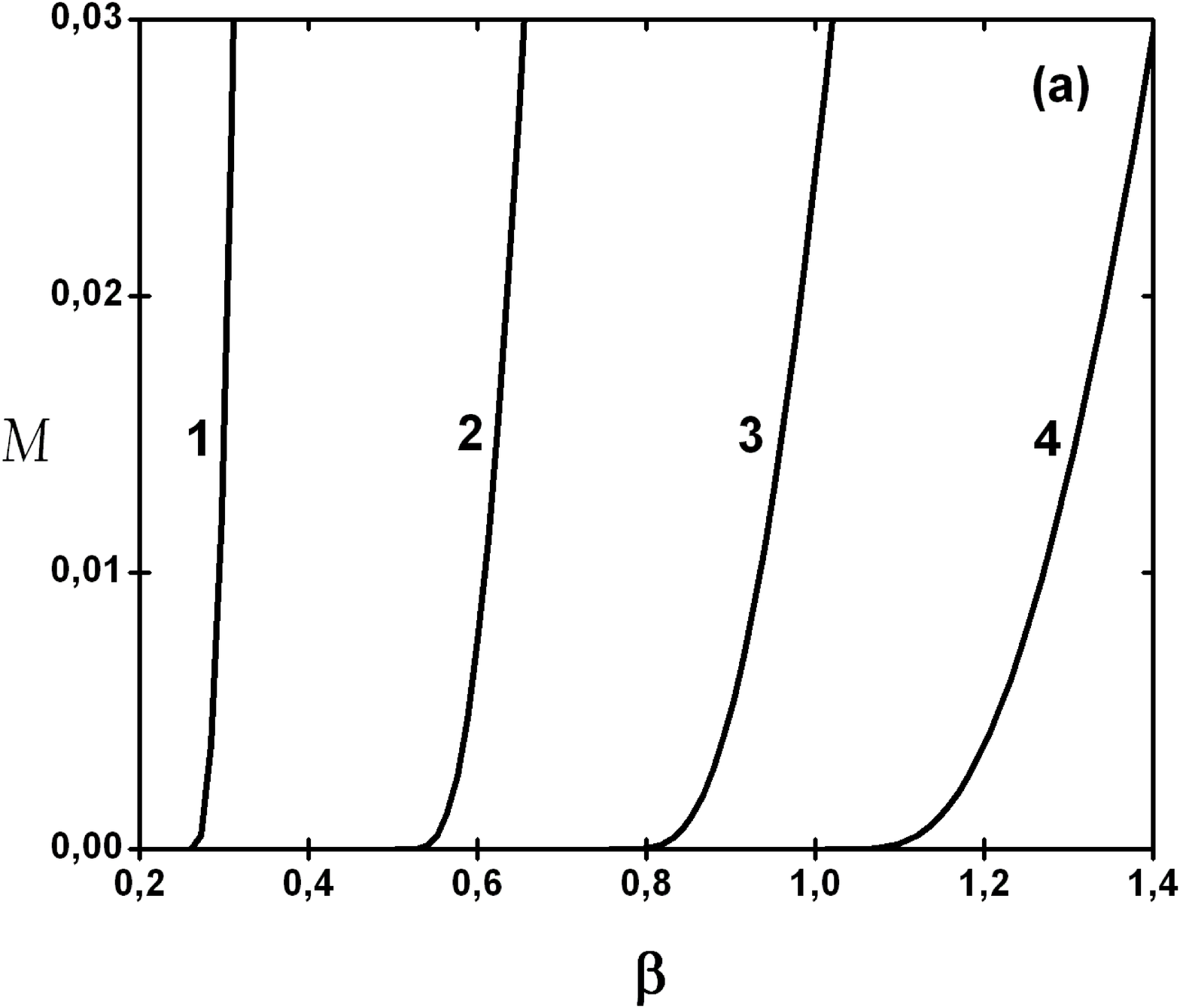}}
\scalebox{0.28}{\includegraphics[angle=0]{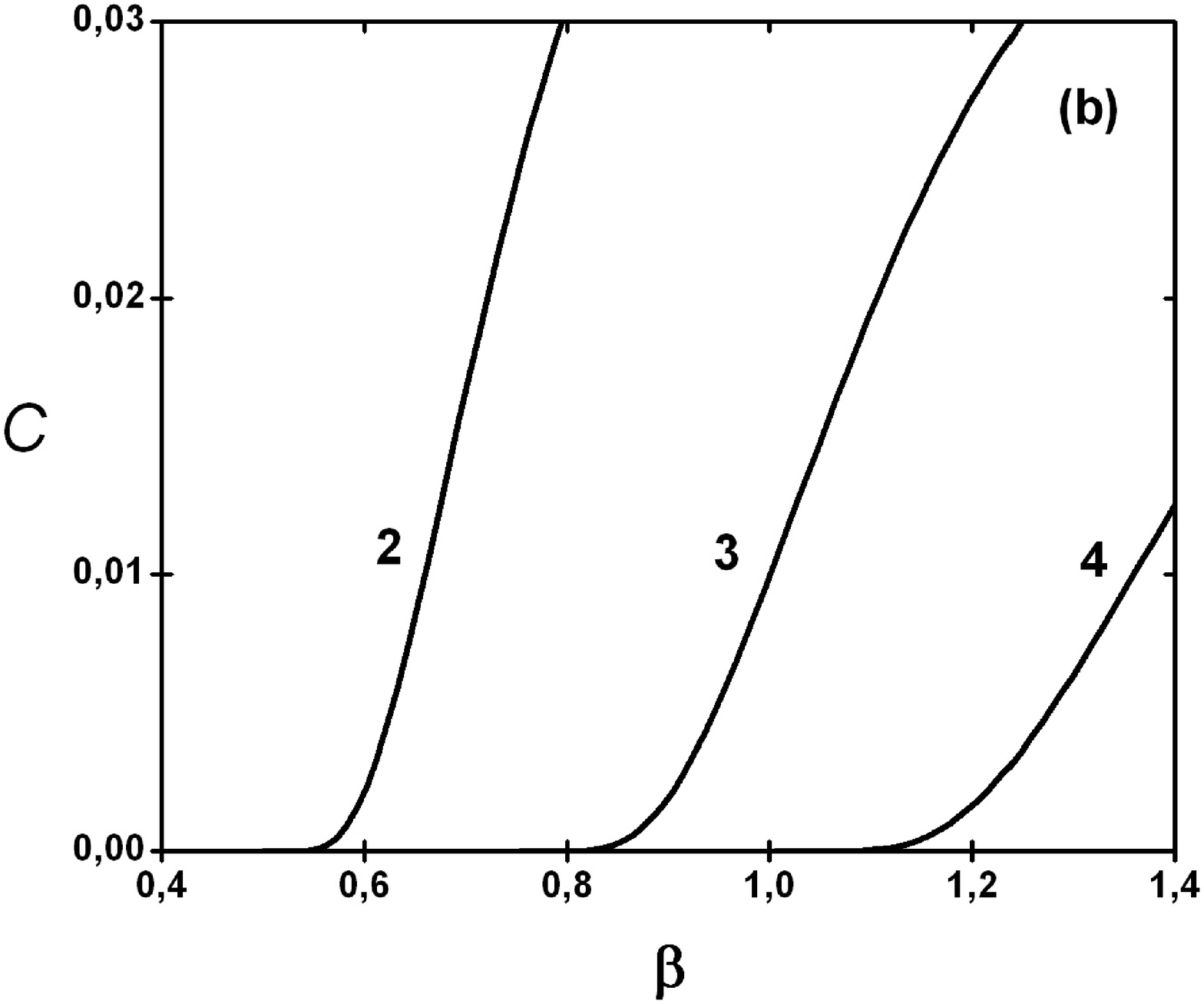}} 
\scalebox{0.18}{\includegraphics[angle=0]{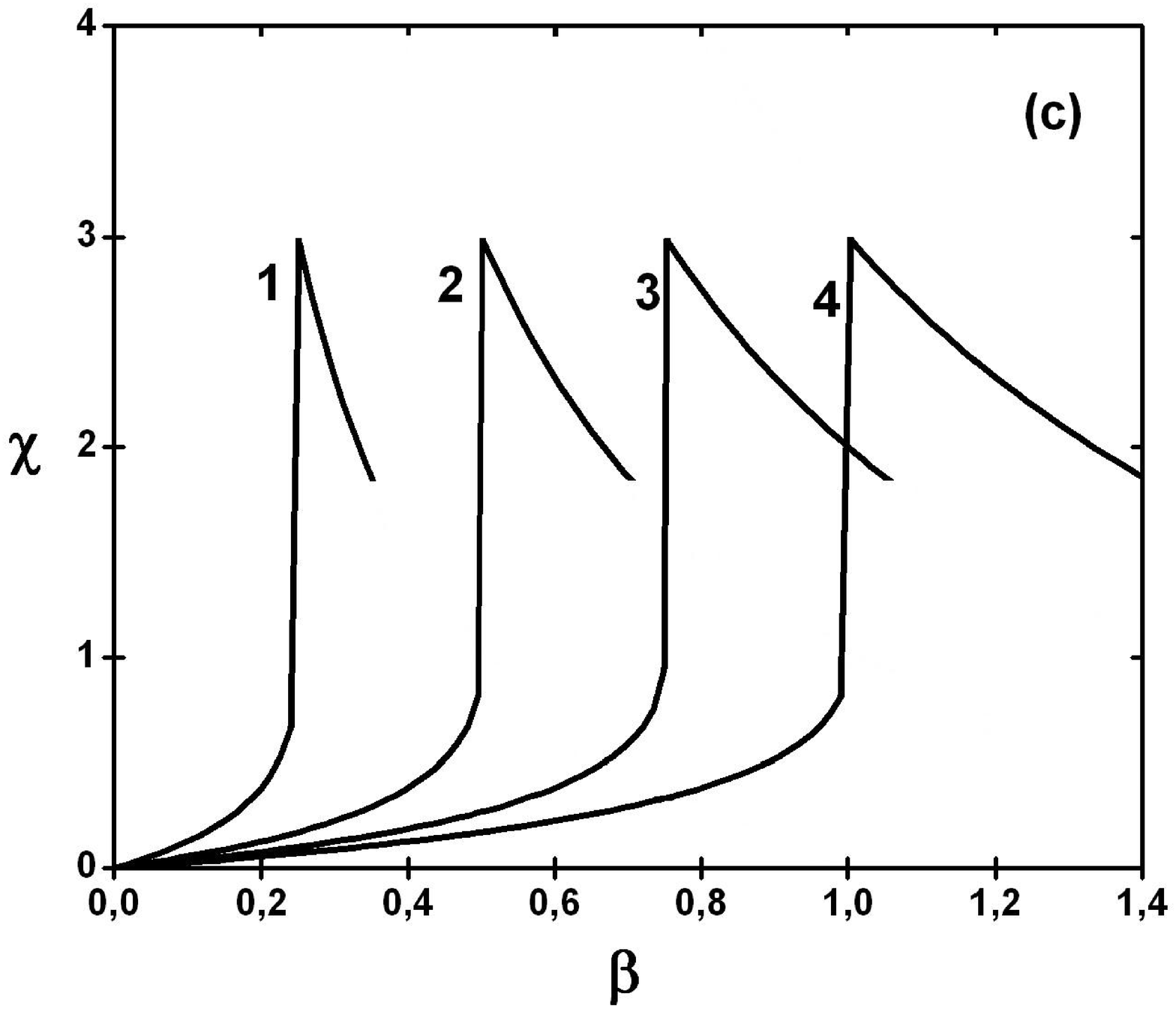}}
\end{center}
\caption{ 
The order parameter $M$ (a), the specific heat $C$ (b), and the susceptibility in zero field, $\chi$, (c) versus inverse temperature $\beta=1/T$ at different $q$ in the $q$-state Potts model on the asymmetric annealed network. The numbers on the curves indicate corresponding $q$. The critical point is $\beta_{c}=1/T_c = q/4$. $\chi(\beta_{c-})=1$, $\chi(\beta_{c+})=3$.
}
\label{58.2}%
\end{figure}

Here we briefly list our results obtained for the infinite annealed network $t\to \infty$.   
We demonstrate that for the all $q \geq 1$, our system has an infinite order phase transition at the critical temperature 
\begin{equation}
T_c = \frac{4}{q}. 
\label{e3}
\end{equation}
In the following we use $\beta \equiv 1/T$ for the sake of convenience. The resulting mean spontaneous ``magnetization'' [$M = q\langle\delta(s,1) -1/q\rangle_T/(q-1)$, where $\langle\cdot\rangle_T$ means the thermodynamic averaging] is  of the typical BKT form 
\begin{equation}
M\left(  \beta\right)  \cong \frac{4ce}{q}e^{\left(  -\pi/\sqrt{4\beta/q
-1}\right)  }  
\label{58}%
\end{equation}
near the critical temperature [see Fig.~\ref{58.2}(a)]. Here $c=1.554\ldots$. Note the difference by factor $2$ of the value $T_{\text{c}}(q{=}2)$ from the corresponding result for the Ising model. This usual difference (see, e.g., Refs.~\cite{dgm04,em05}) disappears after the proper rescaling of the coupling constant in the Hamiltonian (\ref{.5}) to arrive at the Ising Hamiltonian with the coupling constant $1/j$. 

The specific heat $C_{H=0}(T)\equiv C(T)$ per spin is zero in the normal phase, and, below the critical temperature is 
\begin{equation}
C\left(  \beta\right)  \cong \frac{\left(  \pi ce\right)  ^{2}}{128}\frac{q\left(
q-1\right)  }{\left(  \beta-q/4\right)  ^{3}}\exp\left(  -2\pi
/\sqrt{\frac{4}{q}\beta-1}\right)
,   
\label{63}%
\end{equation}
where $(\pi c e)^2/128=1.37\ldots$ [see Fig.~\ref{58.2}(b)]. 

We also find the susceptibility of the system in zero field:       
\begin{eqnarray}
&&
\!\!\!\!\!\!\!\!\!\!\!\!\!\!\!\chi\left(  \beta>q/4\right)  =q/\beta-1 
\ \ \ \ \text{if } \ \beta-q/4 \ll q/4 ,
\nonumber
\\[5pt]
&&
\!\!\!\!\!\!\!\!\!\!\!\!\!\!\!\chi\left(  \beta<q/4\right)  =\left(  1-\sqrt{1-4\beta/q}\right)  /\left(
1+\sqrt{1-4\beta/q}\right) \! . 
\label{63.5}%
\end{eqnarray}
There is a finite jump of the susceptibility at the phase transition
point: $\chi\lbrack\beta{=}(  q/4)_{-}]{=}1$ and $\chi\lbrack
\beta{=}(q/4)_{+}]{=}3$ [see Fig.~\ref{58.2}(c)]. 

The distribution of the linear response $\chi_i$ to a local applied field (i.e., the correlation volume distribution), $P(\mu)  =
\left\langle 
\sum_{i=0}^{t}
\delta\left(  \mu-\chi_{i}\right)\right\rangle/t$, is a power law in the entire normal phase: 
\begin{equation}
P(\mu)  \propto \mu^{-[  1+2/(1-\sqrt{1-4\beta/q})]  }.
\label{67}%
\end{equation}
Below the critical temperature, this characteristic of correlations rapidly decreases. Thus there is a contact of the ``critical'' phase with the power law decreasing distribution and the phase with a fast decrease of $P(\mu)$. At the phase transition point, $P\left(  \mu,\beta=q/4\right)  \propto
\mu^{-3}$.      

Furthermore, we calculate another distribution function 
$P(\nu) = (2/t^2)\left\langle{\textstyle\sum_{i,j}^{t}}
\delta(\nu-\chi_{ij})\right\rangle$, that is the distribution of correlations 
$\sim \chi_{ij}$ in the neighboring spin pairs \cite{remark2} ($\chi_i=\sum_j\chi_{ij}$),  
and above the phase transition, we find a power law with the same exponent: 
\begin{equation}
P(\nu)  \propto \nu^{-[  1+2/(1-\sqrt{1-4\beta/q})]  }
.
\label{68}
\end{equation}

\section{Mean-field equations}\label{mean-field}


It is convenient to rewrite the Hamiltonian (\ref{1}) as 
\begin{equation}
\mathcal{H}=-%
{\displaystyle\sum\limits_{0\leq i<j\leq t}}
\frac{1}{j}
\sum_{p=1}^{q}\delta(s_{i},p)\delta(s_{j},p)
-
{\displaystyle\sum\limits_{i=0}^{t}}
H_{i}\delta(s_{i},1). 
\label{2}%
\end{equation}
Let us assume small fluctuations of $\delta(s_{i},p)$ from their mean-field values $h_{ip}$. In the next section we shall show that this usual (mean-field theory) assumption leads to exact results. 
So, 
one can substitute 
\begin{equation}
\delta(s_{i},p)\delta(s_{j},p) 
\longrightarrow -h_{ip}h_{jp}+h_{ip}\delta
(s_{j},p)+h_{jp}\delta(s_{i},p) 
\label{4}
\end{equation}
into the Hamiltonian~(\ref{2}), 
where $h_{ip}=\left\langle \delta(s_{i},p)\right\rangle$.  
This results in 
a linear effective mean-field Hamiltonian.
The partition function of this Hamiltonian is:
\begin{eqnarray}
&& 
\!\!\!\!\!\!\!\!\!\!\!\!\!\!\!\!\!\!\!\!\!Z=\exp\!\left(-\beta\sum\limits_{p=1}^{q}\sum\limits_{0\leq i<j\leq t}\frac
{h_{ip}h_{jp}}{j}\right)
\times
\nonumber
\\[5pt]
%
&&
\!\!\!\!\!\!\!\!\!\!\!\!\!\!\!\!\!\!\!\!\!{\displaystyle\prod\limits_{i=0}^{t}}\,
{\displaystyle\sum_{p=1}^{q}}
\exp\!\!\left[\beta\!\left(  \sum\limits_{j=0}^{i-1}\frac{h_{jp}}{i}+%
\!\!{\textstyle\sum\limits_{j=i+1}^{t}}
\frac{h_{jp}}{j}\right)\!\!  +\beta H_{i}\delta(p,1)\right]
. 
\label{5.4}%
\end{eqnarray}
For large $t$, one may pass to the continuum limit: $h_{ip}\equiv
h_{p}(x=i/t).$ Then the partition function is 
\begin{eqnarray}
&&
\!\!\!\!\!\!Z=\exp\!\left(-\beta\sum\limits_{p=1}^{q}\int\limits_{0}^{1}\int\limits_{x}%
^{1}\!dxdy\,\frac{h_{p}\left(  x\right)  h_{p}\left(  y\right)  }{y}\right) 
\times
\nonumber
\\[5pt]
&&
\!\!\!\!\!\!\!\!{\displaystyle\prod\limits_{i=0}^{t}}
\sum_{p=1}^{q}\!\exp\!\!\left[\!\beta\!\left( \!\! \frac{1}{x}\!\!\int\limits_{0}^{x}dy\,h_{p}\left(
y\right)\!  + \!\!\!\int\limits_{x}^{1}\!\!dy\,\frac{h_{p}\left(  y\right)  }{y}\!\!\right)\!\!\!
+ \!\beta H(x)  \delta(p,\!1)\!\right]\!, 
\nonumber
\\[5pt]
&&
\label{5.45}%
\end{eqnarray}
and therefore the free energy $F=-\beta^{-1}\ln Z$ is 
\begin{eqnarray}
&& 
\!\!\!\!\!\!\!F    =t\sum_{p=1}^{q}\int\limits_{0}^{1}\int\limits_{x}^{1}dx\ dy\,\frac
{h_{p}\left(  x\right)  h_{p}\left(  y\right)  }{y}-
\nonumber
\\[5pt]
&& 
\!\!\!\!\!\!\!\frac{t}{\beta}\int\limits_{0}^{1}\!dx\,\ln
\left\{  \sum_{p=1}^{q}%
\exp\!\left[\!\beta\!\left(  \frac{1}{x}\int\limits_{0}^{x}\!dy\,h_{p}\left(  y\right)
{+}\int\limits_{x}^{1}\!dy\,\frac{h_{p}\left(  y\right)  }{y}\right)\!  + 
\right.\right.
\nonumber
\\[5pt]
&& 
\left.
\left. 
\ \ \ \ \ \ \ \ \ \ \ \ \ \ \ \ \ \ \ 
\phantom{\int\limits_{0}^{x}}
\beta H(x)  \delta(p,1)
\right]
\right\}  
, 
\label{5.65}%
\end{eqnarray}
which leads to the self-consistent equations for the mean values 
$h_{ip}=\sum_{\left\{  s_{i}%
=1,2...q\right\}  }\delta(s_{i},p)e^{\beta\mathcal{H}\left\{  s_{i}\right\}
}/Z_{i}$ :
\begin{eqnarray}
&& 
\!\!\!\!\!\!\!\!h_{1}\left(  x\right)  = 
\nonumber
\\[5pt]
&&
\!\!\!\!\!\!\!\!\!\frac{1}{A(x)  }\exp\!\left[{\beta\left(  \frac
{1}{x}\int_{0}^{x}\!dy\,h_{1}(y)  +\!\!\int_{x}^{1}\!dy\,\frac{h_{1}(y)  }{y}\right)  {+}\beta H(x) }\right]
,
\nonumber
\\[5pt]
&&
\!\!\!\!\!\!\!\!h_{p}(x)  =  
\nonumber
\\[5pt]
&&
\!\!\!\!\!\!\!\!\!\!\frac{1}{A(x)
}\!\exp\!\left[\beta\!\left(\!  \frac{1}{x}\!\int_{0}^{x}\!\!\!\!dy\,h_{p}(y)  +\!\!\int_{x}%
^{1}\!\!\!\!dy\,\frac{h_{p}(y)  }{y}\!\right) \! \right]\!, \, p=2,...q 
, 
\label{6}%
\end{eqnarray}
where
\begin{eqnarray}
&&
\!\!\!\!\!\!\!A(x)  = 
\nonumber
\\[5pt]
&&
\!\!\!\!\!\!\!\exp\!\left[\beta\left(  \frac{1}{x}\int_{0}^{x}dy\,h_{1}(y)  + \int_{x}^{1}dy\,\frac{h_{1}(y)  }{y}\right)  +\beta
H(x)  \right]+
\nonumber
\\[5pt]
&&
\!\!\!\!\!\!\!\sum_{p=2}^{q}\exp\!\left[\beta\left(  \frac{1}{x}\int_{0}%
^{x}dy\,h_{p}(y)  +\int_{x}^{1}dy\,\frac{h_{p}(y)  }%
{y}\right)  \right]. 
\label{7}
\end{eqnarray}
From Eqs.~(\ref{6}) and (\ref{7}) we obtain  
\begin{eqnarray}
&&
h_{2}\left(  x\right)  =h_{3}\left(  x\right)  =...=h_{q}%
\left(  x\right), 
\nonumber
\\[5pt]
&&
h_{1}\left(  x\right)  +\left(  q-1\right)  h_{p}\left(  x\right)
=1,
\nonumber
\\[5pt]
&&
\!\!\!\!\!\!\!\!\!\!\!\!\!\!\!\!\!\!\frac{h_{1}\left(  x\right)  }{h_{p}\left(  x\right)  }=
\nonumber
\\[5pt]
&&
\!\!\!\!\!\!\!\!\!\!\!\!\!\!\!\!\!\!e^{\beta\left\{
(1/x)\int_{0}^{x}dy\left[ h_{1}(y)  -h_{p}(y)\right]  +\int_{x}^{1}dy
\left[  h_{1}\left(  y\right)-h_{p}\left(  y\right)  \right]/y
\right\}  +\beta H\left(  x\right)  } 
\label{8}%
\end{eqnarray}
for $h_p(x)$. 

The order parameter for the Potts model is defined as 
$
m_i = q\langle\delta(s_i,1) -1/q\rangle_T/(q-1)
$, where $\langle\cdot\rangle_T$ is the thermodynamic average.
So, 
\begin{eqnarray}
&&
h_{1}\left(  x\right)  =\frac{1}{q}\left[  1+\left(  q-1\right)  m\left(
x\right)  \right],  
\nonumber
\\[5pt]
&&
h_{p}\left(  x\right)  =\frac{1}{q}\left[  1-m\left(  x\right)  \right]
, \label{9}%
\end{eqnarray}
These definitions satisfy the first and the second equations of Eqs.~(\ref{8}). 
Substituting Eqs.~(\ref{9}) into the third equation of Eqs.~(\ref{8}) results in the following relation for the order parameter:
\begin{eqnarray}
&&
\!\!\!\!\!\!
m\left(  x\right)  = 
\nonumber
\\[5pt]
&&
\!\!\!\!\!\!\!\!\left\{\exp\left[\beta\!\left(  \frac{1}{x}\int_{0}^{x}\!\!dy\,m\left(
y\right)  +\int_{x}^{1}\!\!dy\,\frac{m(  y)\!  }{y}\right)  +\beta H(x)  \right]\!-\!1\!\right\}/
\nonumber
\\[5pt]
&&
\!\!\!\!\!\!\!\!\left\{\!\exp\!\left[\beta\!\!\left(  \frac{1}{x}\!\int_{0}^{x}\!\!\!\!\!dy\,m(y)
\!+\!\!\int_{x}^{1}\!\!\!\!\!dy\,\frac{m(y)\!\!  }{y}\right)\!\!  +\!\beta H(x)\!
\right]\!{+}(q{-}1)\!  \right\}\!. 
\label{10}%
\end{eqnarray}


\section{Exact free energy}\label{exact}


For simplicity, 
let the external field be homogeneous. The field contribution to the Potts Hamiltonian (\ref{2}) may be written as $-\!\sum_{p=1}^{q}\sum_{i=0}^{t}H_{p}\delta(s_{i},p)$ with $H_p = H\delta(p,1)$. 
Let $\mathcal{H}_{t}$ be the Hamiltonian of the system at time $t$. 
Similarly to the Ising model, see Refs.~\cite{ccg90,cc91,bcd05}, 
one may obtain a recursive relation for $\mathcal{H}_{t}$. This relation leads to  
the following recursive relation for the free energy: 
\begin{equation}
e^{-\beta F_{t}[H_{p}]}= 
\!\!\!\!{\displaystyle\sum\limits_{s_{t}=\left\{  1,2,...q\right\}  }}
e^{-\beta F_{t-1}\left[  H_{p}+
\delta(s_{t},p)/t
\right]  
+\beta
\sum\limits_{p=1}^{q}H_{p}\delta(s_{t},p)}
.
\label{11.4}%
\end{equation}
So, at large $t$, one gets
\begin{equation}
\!\!\!\!\!\!e^{-\beta F_{t}\left(  H_{p}\right)  /t}=%
\!\!\!\!\!{\displaystyle\sum\limits_{s_{t}=\left\{  1,2,...q\right\}  }}\!\!\!\!
e^{\beta\sum_{p=1}^{q}[  H_{p}-(1/t)(\partial F_{t}/\partial
H_{p})]  \delta(s_{t},p)}. 
\label{11.6}%
\end{equation}
Using 
$h_{p}^{(t)}(H_{p})
\equiv\int_{0}^{1}dxh_{p}(x) = -(\partial F_{t}/\partial H_{p})/t$ 
gives 
the 
asymptotically 
exact 
free energy: 
\begin{equation}
F=-\frac{t}{\beta}\ln\sum_{p=1}^{q}\exp\left[\beta\left(h_{p}^{(t)}+H\delta\left(
p,1\right)  \right)  \right]
.  
\label{11.7}%
\end{equation}
On the other hand, this expression may be easily obtained in the framework of the mean-field theory by substitution Eqs.~(\ref{6}) into formula (\ref{5.65}), which demonstrates the exactness of the mean-field theory.   


\section{Solution of Eq.~(\ref{10})} 


Self-consistent equation (\ref{10}) is a direct generalization that for the Ising model, so its solution is straightforward. 
One may see that close to $x=0$, 
\begin{equation}
m\left(  x\sim0\right)  =1-qA(\beta,H)x^{2\beta}, 
\label{12.1}%
\end{equation}
where $A(\beta,H)$ is independent of $x=i/t$. If, however, $H=0$, then $m(x, T{>}T_c)=0$.  
At $x=1$ ($i=t$), we have  
\begin{equation}
m\left(  1\right)  =\frac{e^{\beta\left[  M+H\left(  1\right)  \right]  }%
-1}{e^{\beta\left[  M+H\left(  1\right)  \right]  }+\left(  q-1\right)  }.
\label{12.2}
\end{equation}

It is convenient to define 
\begin{equation}
n\left(  z\right)  =\beta\left(  \frac{1}{x}\int_{0}^{x}dy\,m\left(  y\right)
+\int_{x}^{1}dy\,\frac{m\left(  y\right)  }{y}\right)  +\beta H\left(  x\right)
, 
\label{11.2}%
\end{equation}
where $z=-\ln x$.  
Therefore 
the order parameter in terms of $n(z)$ is 
\begin{equation}
m\left(  x\right)  =\frac{e^{n\left(  z\right)  }-1}{e^{n\left(  z\right)
}+\left(  q-1\right)  }. 
\label{12}%
\end{equation}
If $H=0$, we obtain a second order differential
equation: 
\begin{equation}
\frac{dn\left(  z\right)  }{dz}-\frac{d^{2}n\left(  z\right)  }{dz^{2}}%
=\beta\frac{e^{n\left(  z\right)  }-1}{e^{n\left(  z\right)  }+\left(
q-1\right)  }. 
\label{13.1} 
\end{equation}
The boundary conditions for this equation can be derived from relations (\ref{11.2}) and (\ref{12.1}):  
\begin{eqnarray}
&&\bigskip\left(  \frac{dn\left(  z\right)  }{dz}\right)_{\!\!z=0}=\!n\left(
z=0\right)  
=\beta M 
, 
\nonumber
\\[5pt]
&&
n\left(  z\rightarrow\infty\right)  =\beta z+\text{const}
.
\label{13.2}%
\end{eqnarray}

With variables $n,w(n){\equiv}\beta^{-1}dn/dz$, Eq.~(\ref{13.1}) is reduced to a first order differential equation: 
\begin{equation}
w\frac{dw}{dn}=\beta^{-1}\left(w-\frac{e^{n}-1}{e^{n}+\left(
q-1\right)  }\right)  , 
\label{15}%
\end{equation}
with the boundary conditions 
\begin{eqnarray}%
&&
w\left(  n=\beta M\right)  =M , 
\nonumber
\\[5pt]
&&
w\left(  n\rightarrow\infty\right)  =1
.
\label{49}%
\end{eqnarray}
At small $n$, 
Eq.~(\ref{15}) takes the form 
\begin{equation}
w\frac{d\omega}{dn}=\beta^{-1}\left(w-\frac{n}{q}\right)  .
\label{16}%
\end{equation}
The physical solution of Eq.~(\ref{16}) satisfies   
\begin{eqnarray}
&&
\!\!\!\!\!\!\!\!\!\!\!\!\frac{1}{2}\ln\left[  n^{2}-nqw\left(  n\right)  +\beta qw^2(n)\right]  + 
\nonumber
\\[5pt]
&&
\!\!\!\!\!\!\!\!\!\!\!\!\frac{1}{\sqrt{(4\beta/q)-1}}\arctan\left(
\frac{  2\beta w\left(  n\right)  -n}{n\sqrt{(4\beta/q)-1}}\right)  =C 
,  
\label{20}%
\end{eqnarray}
which gives the 
critical temperature $\beta_{c}=q/4$. Only
for $\beta\geq q/4$ there exists a physical solution. 
Near $T_c$ the solution of Eq.~(\ref{16}) is: 
\begin{equation}
w_{c}\left(  n,\beta_{c}=\frac{q}{4}\right)  =\frac{2n}{q}\left[
1-f\left(  n\right)  \right]  , 
\label{52}%
\end{equation}
where $f\left(  n\right)  $ satisfies the relation 
\begin{equation}
\ln\left[  nf\left(  n\right)  \right]  +\frac{1}{f\left(  n\right)  }=\ln c
. 
\label{53} 
\end{equation}
Here the constant $c=1.554...$ ensures that $w_{c}\left(  n\right)  $ fits the
corresponding solution of Eq.~(\ref{15}) which approaches $1$ as
$n\rightarrow\infty.$ 
To obtain the critical singularity of the order parameter $M(\beta)$, one should, first,  
substitute the boundary condition $w(n=\beta M)  =M$ into relation (\ref{20}). This gives  
\begin{equation}
C=\ln\left(  \beta M\right)  +\frac{1}{\sqrt{(4\beta/q)-1}}%
\arctan\left(  \frac{1}{\sqrt{(4\beta/q)-1}}\right)  
,  
\label{54}%
\end{equation}
i.e., near 
$\beta=q/4$,  
\begin{equation}
C=\ln\left(  \frac{qM}{4}\right)  +\frac{\pi/2}{\sqrt{(4\beta/q)-1}}-1.
\label{55}%
\end{equation}
Second, 
near $\beta=q/4$, in the region $\beta M\ll
n\ll1,$ Eq.~(\ref{20}) takes the form:   
\begin{equation}
\ln\!\left[  n\!\left(  1-\frac{qw(n)  }{2n}\right)  \right]
+\left[  1-\frac{qw(n)  }{2n}\right]  ^{-1}\!\!\!-\frac{\pi
/2}{\sqrt{(4\beta/q){-}1}}=C. 
\label{56}%
\end{equation}
Sewing together, from relations (\ref{52}), (\ref{53}) and (\ref{56}) we have 
\begin{equation}
\ln\left(  c\right)  -\frac{\pi/2}{\sqrt{(4\beta/q)-1}}=C 
,  
\label{57}%
\end{equation}
and so, using 
Eqs.~(\ref{55}) and (\ref{57}), we readily arrive at the result~(\ref{58}) for $M(\beta)$.

Substituting the order parameter from relation (\ref{9}) into Eq.~(\ref{11.7})
we obtain the free energy in terms of the order parameter $M$:  
\begin{equation}
F=-\frac{t}{q}-\frac{t}{\beta}\ln\left[  e^{\beta\left(  q-1\right)
M/q}+\left(  q-1\right)  e^{-\beta M/q}\right]  
. 
\label{60}%
\end{equation}
This leads to the result (\ref{63}) for the specific heat $C(T)=-(T/t)\partial^{2}F/\partial^{2}T$. 
Similarly, one can derive the 
susceptibility 
in zero field, Eq.~(\ref{63.5}). 


\begin{figure}[t]
\epsfxsize=75mm
\begin{center}
\scalebox{0.20}{\includegraphics[angle=0]{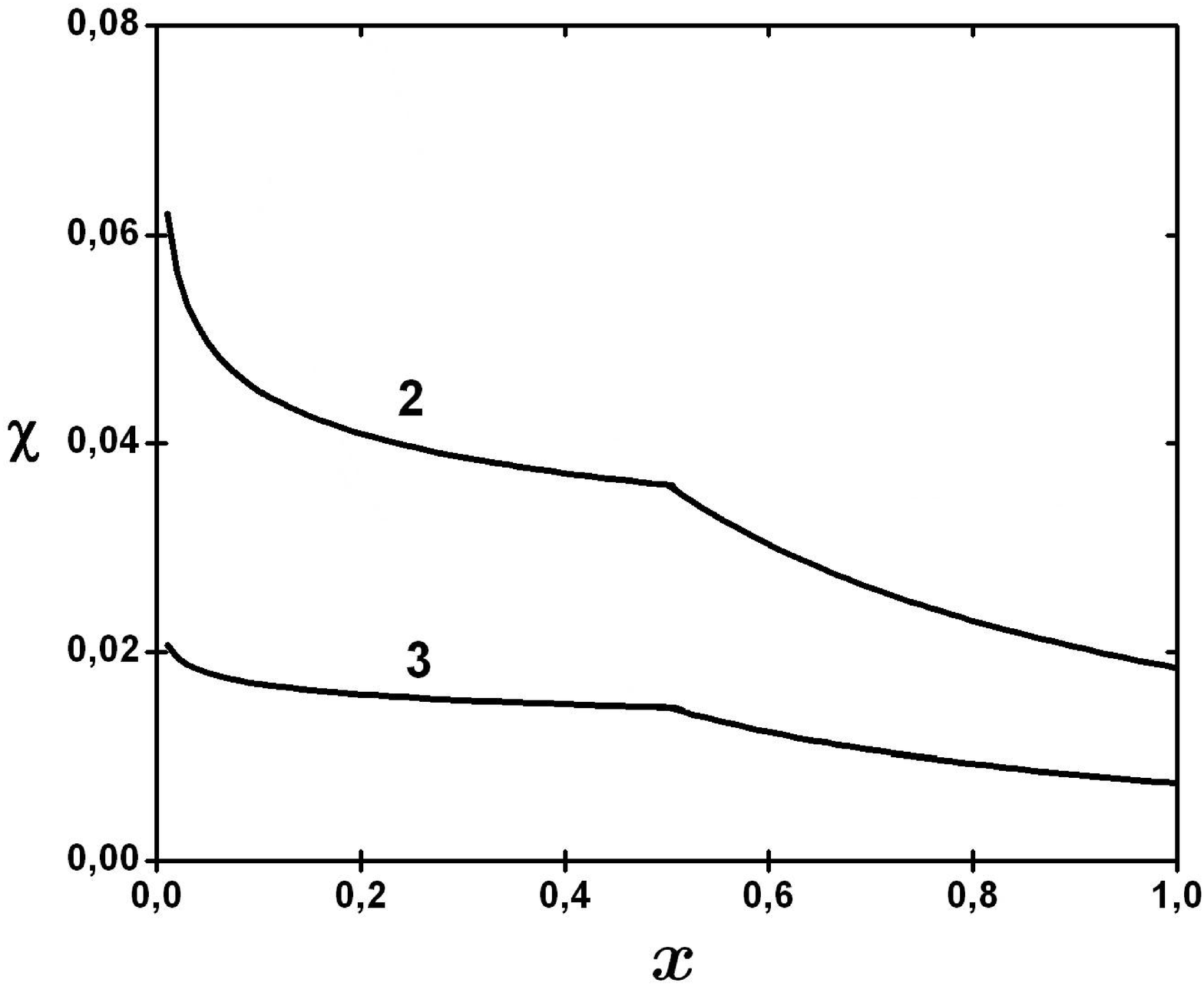}}
\end{center}
\caption{
The local linear response $\chi(x,z)=\partial m(x)/\partial H(z)$ to a field applied at $z=0.5$ for the $2$- and $3$-state Potts models. The values of $q=2,3$ are indicated on curves. The homogeneous component of the applied field is $0$. The inverse temperature $\beta=1/T=0.24>1/T_c(q=2,3)$.   
}
\label{64.1}%
\end{figure}


One may easily find a linear response of the order parameter at point $x$ to a field applied at point $z$: $\chi(x,z)  =\partial m(x)/\partial H(z)$. 
Differentiating of (\ref{10}) by $H\left(  z\right)  $ at $T>T_{c}$ we arrive at the following relation for the linear local susceptibility:
\begin{equation}
\chi\left(  x,z\right)  =\frac{\beta}{q}\!\left[  \frac{1}{x}\!\int_{0}^{x}\!\!%
dy\,\chi\left(  y,z\right)  +\!\!\int_{x}^{1}\!\!\frac{dy}{y}\,\chi\left(  y,z\right)
+\delta\left(  x{-}z\right)  \right]
.  
\label{64}%
\end{equation}
Iterating this equation gives 
\begin{equation}
\!\!\!\!\chi\left(  x,z\right)  =\tilde{\chi}\left(  x,z\right)  +\frac{\beta}%
{q}\,\delta\left(  x-z\right)  
\label{65}%
\end{equation}
(see Fig.~\ref{64.1} for $q=2,3$, and $z=0.5$), 
where%
\begin{eqnarray}
&&
\!\!\!\!\!\!\!\!\!\!\!\!\!\!\tilde{\chi}\left(  x\leq z,z\right)  = 
\frac{\beta^{2}
x^{(-1+\sqrt{1-4\beta/q}\,)/2}
}
{
q^{2}\sqrt{1-4\beta/q}
}
\times
\nonumber
\\[5pt]
&&
\!\!\!\!\!\!\!\!\left\{
z^{(-1-\sqrt{1-4\beta/q}\,)/2  }-\frac{4\beta}{q}\,
\frac{
z^{(-1+\sqrt{1-4\beta/q}\,)/2  }}{\left(  1+\sqrt{1-4\beta/q}\right)
^{2}}\right\}
,
\nonumber
\\[5pt]
&&
\!\!\!\!\!\!\!\!\!\!\!\!\!\!\tilde{\chi}\left(  x>z,z\right)  = \frac{\beta^{2}z^
{(-1+\sqrt{1-4\beta/q}\,)/2}}{q^{2}\sqrt{1-4\beta/q}}\times
\nonumber
\\[5pt]
&&
\!\!\!\!\!\!\!\!\left\{
x^{(-1-\sqrt{1-4\beta/q})/2}-\frac{4\beta}{q}\,
\frac{
x^{(-1+\sqrt{1-4\beta/q})/2}}{\left(  1+\sqrt{1-4\beta/q}\right)
^{2}}\right\}
\label{65a}%
\end{eqnarray}
(compare with Refs.~\cite{ccg90,cc91}). 
So the total linear response to a local field applied at $z$, 
$\chi(z)  =\int_{0}^{1}dx\chi(x,z)$, is:
\begin{equation}
\chi(z)= 
\frac{2\beta/q}{1+\sqrt{1-4\beta/q}
}\,z^{-(1-\sqrt{1-4\beta/q})/2}
. 
\label{66}%
\end{equation}
These formulas directly lead to results (\ref{67}) and (\ref{68}) in Sec.~\ref{s-results} for the distributions of the ``correlation volume'' and of $\chi_{ij}$.


\section{Discussion and summary}


We have found that for any number of states $q$, the critical behavior of the $q$-state Potts model on the asymmetric annealed network is quite similar to that of the Ising model on this substrate, i.e., of the $2$-state Potts model. 
Moreover, no qualitative difference with the Ising model was observed at any $q$. 
This is despite the contrasting behaviors of the $q{<}3$- and $q{\geq}3$-Potts models in the traditional mean-field theory. The fact that a model with a first order phase transition in the usual mean-field regime may show the infinite order phase phase transition with BKT critical singularity, if it is placed on the specific asymmetric annealed network, is the main result of this paper. 

Above $T_c(q)$, the correlation volume distribution $P(\mu)$ is power-law, i.e., critical. Thus $T_c(q)$ separates the ``critical phase'' and the phase [in this case $T>T_c(q)$] where this distribution rapidly decreases. This is a generic feature of the BKT transitions. Interestingly, we have found that the distribution $P(\nu)$ of the pairwise correlations---the distribution of the $\chi_{ij}$ value---asymptotically coincides with $P(\mu)$ above $T_c$, which indicates that large $\chi_i=\sum_j \chi_{ij}$ responses (``correlation volumes'') are determined only by large $\chi_{ij}$ contributions. A power-law $P(\nu)$ distribution was earlier found in the entire normal phase of a different cooperative model on a quite different network substrate, see Ref.~\cite{mr05}.        

Our results, generalizing the earlier findings for the Ising model, were obtained for a very specific annealed network. We suggest that resembling critical features may be found in the $q$-state Potts model on the corresponding growing networks with ``quenched'' disorder and an exponentially decreasing degree distribution. Our findings show that the BKT critical singularity should occur in a wider range of cooperative systems that one might expect.


\begin{acknowledgments} 

This work was partially supported by projects 
POCTI/FAT/46241/2002, POCTI/MAT/46176/2002, and DYSONET. 
Authors thank A.~V.~Goltsev and A.~N.~Samukhin for useful discussions. SND acknowledges M.~Bauer and S.~Coulomb (CEA-Saclay) for numerous ideas proposed when studying the BKT-like transition in the Ising model on a network.

\end{acknowledgments}


\begin{thebibliography}{9}
\bibitem{remark}
One may see, however, that the annealed network model cannot distinguish a tree (i.e., a graph without loops) from a loopy network. In particular, in this model, we could make, instead of one, say, two connections between each vertex and its predecessors, and arrive, after rescaling the coupling $1\to1/2$, at the same system.
\bibitem{remark2} 
In fact, in this system, each two spins are the nearest neighbors.
\end{thebibliography}
\end{document}